\documentclass[aps,prl,twocolumn,showpacs,superscriptaddress,groupedaddress]{revtex4}  
\usepackage{graphicx}  
\usepackage{dcolumn}   
\usepackage{bm}        
\usepackage{amssymb}   

\newcommand{\be}{\begin{equation}}
\newcommand{\ee}{\end{equation}}
\newcommand{\bea}{\begin{eqnarray}}
\newcommand{\eea}{\end{eqnarray}}

\newcommand{\rme}{{\rm{e}}}

\begin{document}

\title{The double-padlock problem: is secure classical information transmission possible without key exchange?}  

\author{James M.~Chappell}
\email{james.m.chappell@adelaide.edu.au}
\affiliation{School of Electrical and Electronic Engineering, University of Adelaide,
SA 5005, Australia}
\author{Derek Abbott}
\affiliation{School of Electrical and Electronic Engineering, University of Adelaide,
SA 5005, Australia}


\date{\today}

\begin{abstract}
The idealized Kish-Sethuraman (KS) cipher is theoretically known to offer perfect security through a classical information channel. However, realization of the protocol is hitherto an open problem, as the required mathematical operators have not been identified in the previous literature. A mechanical analogy of this protocol can be seen as sending a message in a box using two padlocks; one locked by the Sender and the other locked by the Receiver, so that theoretically the message remains secure at all times. We seek a mathematical representation of this process, considering that it would be very unusual if there was a physical process with no mathematical description and indeed we find a solution within a four dimensional Clifford algebra. The significance of finding a mathematical description that describes the protocol, is that it is a possible step toward a physical realization having benefits in increased security with reduced complexity.
\end{abstract}

\maketitle

Various schemes exist to maintain secure information channels that exploit  physical phenomena such as quantum effects \cite{Buhrman2012complete,lo2012measurement} (eg.~indeterminacy, entanglement) or even classical chaos \cite{lo2012measurement,nguimdo2011digital,kanter2008public}.  All existing schemes involve, one way or another, the sharing or exchange of a cryptographic key.  The open question we address  in this paper is: can secure transmission  be achieved without any form of key exchange? And if so, which physical property of nature can be exploited to achieve this?

The {\it Kish-Sethuraman cipher} (KS-cipher) is an idealized protocol that achieves the goal of avoiding key exchange \cite{laszlo2004non,kish2005,klappenecker2004remark}. However, this protocol has not yet been realized, as the appropriate physical property, with a supporting mathematical description, has not yet been identified. In this paper we show that classical operations on a Clifford space remarkably possess the required mathematical properties and we develop an appropriate ansatz based on Clifford algebra.

\begin{figure}[htb]

\begin{center}
\includegraphics[width=3.4in]{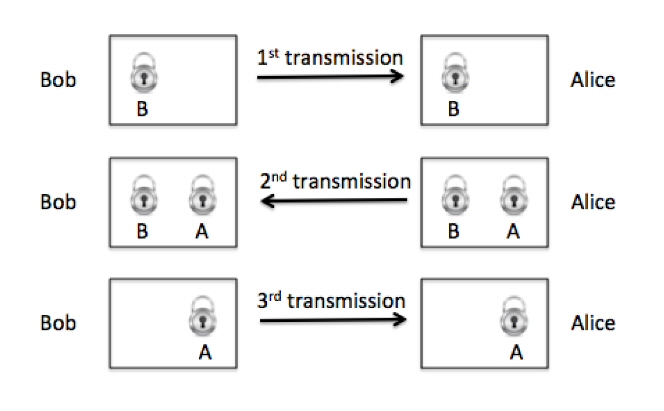}
\end{center}

\caption{The double padlock protocol of Kish and Sethuraman.  Bob firstly locks the box and sends it to Alice.  Then, once received, Alice also padlocks the box with a second lock and sends it back to Bob.  Finally, Bob unlocks his padlock, and sends the box back to Alice who can then remove her lock, open the box, and read the message.  The message appears perfectly secure because at all times it has been secured by at least one lock.
\label{doublepadlock}}
\end{figure}

First, let us briefly review how the Kish-Sethuraman cipher protocol works, using a mechanical analogy. Suppose Bob wishes to transmit a written message to Alice; Bob hides the message in a box that he securely padlocks before sending it to Alice.  After receiving the box, Alice adds a second padlock and sends the box back to Bob.   Then Bob unlocks his padlock, leaving the box still secured by Alice's lock, and sends it back to Alice who can then remove her lock, open the box and read the message as shown in Fig.~\ref{doublepadlock}.

This KS-cipher protocol is perfectly secure because both Bob and Alice keep their keys  undisclosed so that at all times the box is locked by at least one padlock, thus no information is leaked or shared \cite{kish2005}.
Hence we can say that in the physical world, a completely secure classical protocol is conceptually possible.  In practice, a physical box can be broken, however, what is important to our analysis is the security of the lock protocol. This physical example is clearly classical and so we would expect that there would be a mathematical model to describe this process. That is, it would seem strange if there was such a simple physical scenario  for which there was no counterpart in the mathematical world and so would run counter to general trend of the  success of mathematics in describing the physical world. This then underlies the motivation for expecting that a mathematical description might indeed be feasible.

The significance of a mathematical protocol simulating the double-padlock problem is that it would potentially be the underpinnings of a relatively simple method of avoiding key exchange for secure information transmission.

Firstly we note that the ordering of the padlocks commutes.  That is Alice and Bob can take off or add their padlock in any order, which is the primary aspect of the protocol that permits it to work and hence we are looking to find two mathematical operations that can be applied by Alice and Bob that commute.  We can immediately identify an example of this in the case of two-dimensional rotations.

For example, the message Bob wants to secretly send could be the value $ \theta $.
Bob `hides' $ \theta $ by adding a random angle $ \phi_1 $ (his `key') to it and sends it to Alice.  
Then Alice adds another random angle $ \phi_2 $ (her `key') and sends it back to Bob. 
Then Bob undoes his secret rotation $ \phi_1 $ and returns the message to Alice.  
Then Alice undoes her rotation $ \phi_2 $ and recovers the original value of $ \theta $.
These operations are most elegantly analyzed in two-dimensional geometric algebra, where we have a message vector  $  \mathbf{m} = m_1 e_1 + m_2 e_2 $, using $ e_1 $ and $ e_2 $ as orthogonal basis elements and producing the bivector iota $ \iota = e_1 e_2 $.  Acting on the message vector with a rotor $ R = \rme^{ \iota \phi/2 } $ produces a rotated vector
\be
 \mathbf{m}' = R \mathbf{m} \tilde{R} = \rme^{ \iota \phi/2 } \mathbf{m} \rme^{ - \iota \phi/2 } ,
\ee
where $  \mathbf{m}' = m_1' e_1 + m_2' e_2 $ and where we have defined the {\it reversion} operation, which inverts the order of all algebraic products, that is, $ \tilde{R} = \rme^{ -\iota \phi/2 } $.  Therefore $ \phi $ in this case represents the private key and rotates the vector $ \mathbf{m} $ by a clockwise angle $ \phi $. In two dimensions, we can combine the two sides of the rotation operator because $ \iota $ anticommutes with both $ e_1 $ and $ e_2 $ within the vector $  \mathbf{m}$, so that $ \mathbf{m}' = \rme^{ \iota \phi } \mathbf{m} $.  Refer to the Appendix for a brief summary of these operations that utilize geometric algebra.
Therefore, after the operations by Alice and Bob we find
\be
 \mathbf{m}_{\rm{final}} = \tilde{R}_A \tilde{R}_B R_A R_B  \mathbf{m} = \tilde{R}_A R_A \tilde{R}_B  R_B  \mathbf{m} = \mathbf{m} ,
\ee
where because the rotation operators commute and $ \tilde{R}_A R_A  =  \tilde{R}_B R_B = 1 $, we recover the initial message.  The message (the angle with the $ e_1 $ axis say) can be recovered from $ \cos \theta =  \mathbf{m} \cdot e_1 / |  \mathbf{m} | $, where the vector length $ |  \mathbf{m} | = \sqrt{\mathbf{m}^2} $.

While this process indeed hides the message at each stage, an eavesdropper, Eve, by comparing the successive intermediate transmissions, can deduce the intermediate rotations and hence discover the two keys ($\phi_1 $ and $ \phi_2 $) thereby unlocking the message.  That is, intercepting two consecutive transmissions, which consist of two-dimensional vectors, Eve can easily calculate the rotation angle between them from $ \mathbf{m}_2 = \rme^{ \iota \phi } \mathbf{m}_1 $, which can be rearranged to give $  \rme^{ \iota \phi } = \mathbf{m}_2 \mathbf{m}_1^{-1} $.  The inverse of a vector being easily calculated when it is represented in geometric algebra, as shown in the Appendix.

In an attempt to circumvent the vulnerability of two-dimensional rotations, we can consider more general operators using two-dimensional multivectors
\be
M = a +\mathbf{v} + \iota b ,
\ee
where $ a $ and $ b $ are scalars, $ \iota $ is the bivector and a planar vector $ \mathbf{v} = v_1 e_1 + v_2 e_2 $.  That is $ \bigwedge \Re^2  $ is the exterior algebra of $ \Re^2 $ which produces the space of multivectors $ \Re \oplus \Re^2 \oplus \bigwedge^2 \Re^2 $, a four-dimensional real vector space denoted by $ Cl_{2,0}(\Re) $.
We now have the encryption process
\be \label{multivectorOperators}
\mathbf{m}_{\rm{final}}  = M_A^{\dagger} M_B^{\dagger}  M_A M_B  \mathbf{m} M_B^{\dagger} M_A^{\dagger} M_B M_A  ,
\ee
where the $ \dagger $ operation is an {\it inverse} operation, not necessarily the reversion operation, such that $ M_A^{\dagger} M_A  = M_B^{\dagger} M_B  = 1 $.  The first message sent by Bob to Alice is then $ \mathbf{m}_1 = M_B  \mathbf{m} M_B^{\dagger} $, who then returns $ \mathbf{m}_2 = M_A M_B  \mathbf{m} M_B^{\dagger} M_A^{\dagger} $ which Bob then sends back to Alice as $ \mathbf{m}_3 = M_B^{\dagger}  M_A M_B  \mathbf{m} M_B^{\dagger} M_A^{\dagger} M_B $, who can then decode the message as shown in Eq.~(\ref{multivectorOperators}).

So, seeking commuting operators $ M_A $ and $ M_B $, that is $ M_A M_B - M_B M_A = 0 $ we require
\bea \nonumber
& & \left ( a+ \mathbf{v} + \iota b \right )\left ( c+ \mathbf{w} + \iota d \right ) - \left ( c+ \mathbf{w} + \iota d \right ) \left ( a+ \mathbf{v} + \iota b \right ) \\ 
& & = 2 \mathbf{v} \wedge \mathbf{w} - 2 \iota d \mathbf{v} + 2 \iota b \mathbf{w} = 0  .
\eea
We therefore require $ \mathbf{v} $ and $ \mathbf{w} $ to be parallel, and so we need to select a preferred direction for the protocol during handshaking, say the direction $ e_1 $.  Hence Alice and Bob can utilize multivectors
\be
M_A = a + v e_1 + \iota v \, , \, M_B = b + w e_1 +  \iota w 
\ee
that when normalized can be written as $ M_A = \rme^{v e_1 + \iota v} $ and  $ M_B = \rme^{w e_1 + \iota w} $, and defining $ M_A^{\dagger} = \rme^{-v e_1 - \iota v} $ and $ M_B^{\dagger} = \rme^{-w e_1 - \iota w} $, we have $ M_A M_A^{\dagger} = M_B M_B^{\dagger} =1 $.  The one degree of freedom in the operator is insufficient to ensure security of the two-dimensional message vector and so we need to seek a solution in higher dimensions.  

In three dimensions, we have a message vector $ \mathbf{m} = m_1 e_1 + m_2 e_2 + m_3 e_3 $ and define the trivector $ i = e_1 e_2 e_3 $ that commutes with all variables with $ i^2 = (e_1 e_2 e_3 )^2 = -1 $. 

We can write general three-dimensional multivector operators for Alice and Bob as 
\be
M_A =  a+ \mathbf{v} + i \mathbf{r} + i b  \, , \, M_B =  c+ \mathbf{w} + i \mathbf{s} + i d
\ee
where $ \mathbf{v} $ and $ \mathbf{r} $ are three-vectors. This is the space of multivectors $ \Re \oplus \Re^3 \oplus \bigwedge^2 \Re^3 \oplus \bigwedge^3 \Re^3 $, an eight-dimensional real vector space denoted by $ Cl_{3,0}(\Re) $.
We now seek $ M_A $ and $ M_B $ to be commuting in order to use the procedure in Eq.~(\ref{multivectorOperators}), requiring
\bea
0 & = & M_A M_B - M_B M_A  \\ \nonumber
& = & 2 ( \mathbf{v} \wedge  \mathbf{w} - \mathbf{r} \wedge  \mathbf{s} ) + 2 i ( \mathbf{v} \wedge  \mathbf{s} + \mathbf{r} \wedge  \mathbf{w} ) ,
\eea
and to make this commutator vanish we can select $ \mathbf{w} = \mathbf{v} \iota =  \mathbf{v} i e_3 $ and $ \mathbf{s} = \mathbf{r} \iota = \mathbf{r} i e_3  $, with the vectors now planar in order to anticommute with $ e_3 $, so we define $ \mathbf{v}_{12} = v_1 e_1 + v_2 e_2 $. We could have selected a general direction, in place of the direction $ e_3 $, however this direction needs to be shared publicly, and so without loss of generality we can select the $ e_3 $ direction. That is, we have the commuting operators
\bea
M_A & = & \left ( a + \mathbf{v}_{12} + e_3 \mathbf{v}_{12}  + i b \right ) = \rme^{ i \phi_1} \rme^{ (1+ e_3 ) \mathbf{x}_{12} } \\ \nonumber
M_B & = & \left ( c + \mathbf{w}_{12} + e_3 \mathbf{w}_{12} + i d \right ) = \rme^{ i \phi_2} \rme^{(1 + e_3 ) \mathbf{y}_{12} }, \nonumber
\eea
which we have written in an exponential form to guarantee normalization with planar vectors $ \mathbf{x}_{12} $ and $ \mathbf{y}_{12} $,
with the encrypted message for Alice, for example, given by
\be \label{3Dencrypt}
\mathbf{m}' = M_A \mathbf{m} M_A^{\dagger} .
\ee
However, we can see that the leading phase term in the operator commutes through the message vector $ \mathbf{m} $ and so leaves only two degrees of freedom available to encrypt the message, insufficient to stop an eavesdropper.  

This can also be understood through the example of general three dimensional rotations.  In this case we rotate a unit vector (with two degrees of freedom) through an action by the rotor (consisting of a rotation axis with two degrees of freedom) and a rotation angle giving a total of three degrees of freedom.  We can see that with the knowledge of the start and final vectors, we can {\it not} determine the full details of the rotor.  However in three dimensions rotations do not commute and so it appears that we need to implement some form of rotation within a higher four dimensional space.

Hence, we need to explore if the scheme can work in four dimensions.
In four dimensions we have the space of multivectors $ \Re \oplus \Re^4 \oplus \bigwedge^2 \Re^4 \oplus \bigwedge^3 \Re^3 \oplus \bigwedge^4 \Re^4$, a sixteen-dimensional real vector space denoted by $ Cl_{4,0}(\Re) $. We select a message four-vector $ \mathbf{m} = m_1 e_1 + m_2 e_2 + m_3 e_3 + m_4 e_4 $ and we define the quadvector $ I = e_1 e_2 e_3 e_4 $ that anticommutes with all vectors and has a positive square. 
Now, requiring $ M_A M_B = M_B M_A $, after some algebra detailed in the Appendix, we find four types of commuting multivectors, and the type describing pure rotations in four dimensions produce the commuting operators 
\be
M_A =  a + e_4 ( \mathbf{v} - I \mathbf{v})  \, , \, M_B =  b + e_4 ( \mathbf{p} + I \mathbf{p}) ,
\ee
where $ \mathbf{v}, \mathbf{p} $ are four-vectors. We thus have four degrees of freedom for the private keys for both Alice and Bob respectively, as the values of $ a $ and $ b $ are fixed by the requirement of normalization.  In order to reveal more clearly that these operators lie on the even subalgebra, we can write the operator for Alice, for example, as
\be
M_A =  (a + v_4) + e_4 \vec{v} + i \vec{v} + I v_4 ,
\ee
where $ \mathbf{v} = \vec{v} + v_4 e_4 $.  Because the operators lie on the even subalgebra we can encrypt the messages using the reversion operation, with 
\be
 \mathbf{m}' = M \mathbf{m} \tilde{M} ,
\ee
which maps from unit four-vectors to unit four-vectors. 
Hence Eve needs to discover the private key $ \mathbf{v}$ with four degrees of freedom, whereas $ \mathbf{m}' $ and $ \mathbf{m} $ are the intercepted intermediate message unit four-vectors having three degrees of freedom.  Hence we find a similar situation to that found for rotations in three dimensions discussed earlier, where the rotation axis cannot be determined given the initial and final vectors, but this time in four dimensions with an unknown rotation plane. 

In this paper, for the first time, we provide a set of working mathematical operators for the Kish-Sethuraman (KS) cipher that is a classically secure protocol. Our solution requires the use of the space of Clifford multivectors, we find a viable solution in four dimensional space, and future exploration in dimensions higher than four may be of fundamental interest. 

The encoding of these multidimensional operations onto real signals remains an open question for further study, and it is worth noting that various multidimensional spaces are already exploited by engineers in standard communications theory, for example see \cite{Hajjar2009steered}.

Whilst it is of interest for future work to explore how to physically encode higher dimensional rotations on a wireless carrier signal, the scheme we have developed has wider implications. For example, Klappenecker has conjectured a connection between a mathematical realization of the KS-cipher protocol and the P versus NP problem in computer science \cite{klappenecker2004remark}. Thus it may be of interest to explore implications of the KS operations developed in this paper on the P versus NP problem.

If our mathematical protocol can be encoded on a wireless carrier or fiber optic signal, a benefit would be secure communication without key exchange and the promise of a relatively simple physical realization.

\section{appendix}

\subsection{Geometric algebra representation of vectors}

In order to represent the three independent degrees of freedom of space, Clifford defined an associative algebra consisting of three elements $ e_1 $, $ e_2 $ and $ e_3 $, with the properties
\be \label{orthonormality}
e_1^2 = e_2^2  = e_3^2 =  1 
\ee
but with each element anticommuting, that is $ e_j e_k = - e_k e_j $, for $ j \ne k $. We also define the trivector $ i = e_1 e_2 e_3 $, which allows us to write $ e_2 e_3 = i e_1 $,  $ e_3 e_1 = i e_2 $ and $ e_1 e_2 = i e_3 $.

Now, given two vectors $ \mathbf{a} = a_1 e_1 + a_2 e_2 + a_3 e_3 $ and $ \mathbf{b} = b_1 e_1 + b_2 e_2 + b_3 e_3 $, using the distributive law for multiplication over addition \cite{Doran2003}, as assumed for an algebraic field, we find their product
\bea \label{VectorProductExpand2DInitial}
\textbf{a} \textbf{b} & = & (a_1 e_1 + a_2 e_2 + a_3 e_3)( b_1 e_1 + b_2 e_2 + b_3 e_3) \\ \nonumber
& = & a_1 b_1 + a_2 b_2 + a_3 b_3 +  (a_2 b_3 - a_3 b_2 ) e_2 e_3  \\ \nonumber
& & + (a_3 b_1 - a_1 b_3 ) e_3 e_1 + (a_1 b_2 - a_2 b_1 ) e_1 e_2  ,
\eea
where we have used the elementary properties of $ e_1,e_2,e_3 $ defined in Eq.~(\ref{orthonormality}).  Recognizing the dot and wedge products, we can write
\be \label{VectorProductExpand2D}
\textbf{a} \textbf{b} = \textbf{a} \cdot \textbf{b}  + \textbf{a} \wedge \textbf{b} .
\ee
We can see from Eq.~(\ref{VectorProductExpand2DInitial}), that the square of a vector $ \mathbf{a}^2 = \mathbf{a} \cdot \mathbf{a} = a_1^2 + a_2^2 + a_3^2 $, becomes a scalar quantity.
Hence the Pythagorean length of a vector is simply $ |\mathbf{a}| = \sqrt{\mathbf{a}^2} $, and so we can find the inverse vector 
\be \label{inversevector}
\mathbf{a}^{-1} =  \frac{\mathbf{a}}{\mathbf{a}^2}.
\ee
These results can easily be adapted for a space of any number of dimensions.

\subsection{Derivation of commuting operators in 4D}

We can write a general multivector in four dimensions as 
\be \label{4Dmultivector}
M_A = \mathbf{v} + I \mathbf{w} + e_4 \left ( \mathbf{x} + I \mathbf{y} \right ) = x_4 + \mathbf{v} + e_4 \vec{x} - i \vec{y} + I \mathbf{w} - y_4 I 
\ee
thus forming the complete set of scalar, vector, bivector, trivector and quadvector components, where $ \vec{x} $ and $ \vec{y} $ are three-vectors and $ \mathbf{v} $, $ \mathbf{w} $, $ \mathbf{x} $, $ \mathbf{y} $ are four vectors. We also define similarly $ M_B =  \mathbf{p} + I \mathbf{q} + e_4 \left ( \mathbf{r} + I \mathbf{s} \right ) $.

For two four dimensional multivector operators $ M_A $ and $ M_B $ we have the grade selected by brackets $ \langle  \rangle_g $, where $ g $ is the multivector grade.  Defining the commutator as $ C = M_A M_B -M_B M_A $, we find
\bea
\langle C \rangle_0  & = &  0 \\ \nonumber 
\langle C \rangle_1 & = & - \vec{x} p_4 + v_4 \vec{r} - i \vec{v} \wedge \vec{s} - i \vec{y} \wedge \vec{p}  \\ \nonumber
& & + e_4 ( \vec{x} \cdot \vec{p} -\vec{v} \cdot \vec{r}  + w_4 s_4 - y_4 q_4 ) \\ \nonumber 
\langle C \rangle_2 & = &  2 ( \mathbf{v} \wedge \mathbf{p} - \mathbf{w} \wedge \mathbf{q} - \vec{x} \wedge \vec{r}  - \vec{y} \wedge \vec{s} \\ \nonumber
& &  + I ( \vec{x} \wedge \vec{s} + \vec{y} \wedge \vec{r}) ) \\ \nonumber
\langle C \rangle_3 & = & I \vec{v} s_4 - I \vec{s} v_4 - i \vec{w} \cdot \vec{r} + I \vec{r} w_4 - e_4 \vec{w} \wedge \vec{s} + i \vec{x} \cdot \vec{q} \\ \nonumber 
& & - I \vec{x} q_4 + I \vec{y} p_4 - I \vec{p} y_4 -e_4 \vec{y} \wedge \vec{q}  \\ \nonumber
\langle C \rangle_4 & = &  2 I ( \mathbf{w} \cdot \mathbf{p} - \mathbf{v} \cdot \mathbf{q} ). \nonumber
\eea
By inspection of the quadvector and bivector terms we identify a solution $ \mathbf{v} = \pm \mathbf{w} $ and $ \mathbf{p} = \pm \mathbf{q} $ with the condition $ - \vec{x} \wedge \vec{r} - \vec{y} \wedge \vec{s} = 0 $ and $ I ( \vec{x} \wedge \vec{s} + \vec{y} \wedge \vec{r} ) = 0 $ that implies $ \vec{x} = \pm \vec{y} $ and $ \vec{r} = \mp \vec{s} $. We then will find that the vector and trivector conditions are satisfied as well provided $ \mathbf{x} = - \mathbf{v}' $ and $ \mathbf{r} =  \mathbf{p} $, where $ \mathbf{v}' = e_4 \mathbf{v} e_4 = -v_1 e_1 -v_2 e_2 - v_3 e_3 + v_4 e_4 $.  This then gives two commuting multivectors
\bea \label{nonzerocommute} \nonumber
M_A & = & a + \mathbf{v} + I \mathbf{v} -( \mathbf{v} + I \mathbf{v} ) e_4 = a + ( \mathbf{v} + I \mathbf{v} ) (1- e_4 ) \\ \nonumber
M_B & = & c + \mathbf{p} + I \mathbf{p} + e_4 ( \mathbf{p} + I \mathbf{p} ) =c + (1 + e_4 )( \mathbf{p} + I \mathbf{p} ) .
\eea
From the bivector condition, we could have selected the alternative $ \mathbf{x} = \mathbf{y} = 0 $, that also leads to commuting multivectors 
\be \label{appCommute1}
M_A = a + \mathbf{v} + I \mathbf{v} \, , \, M_B = c + \mathbf{p} + I \mathbf{p} .
\ee
A third type can be found as
\be
M_A = 1 + (1+e_4) (\vec{v} + s I) \, , \, M_B = 1 + (1+e_4) (\vec{w} + t I).
\ee
Alternatively selecting $ \mathbf{q} = \mathbf{w} = 0 $ from the quadvector condition, we find the commuting operators
\be \label{appCommute2}
M_A =  b + e_4 ( \mathbf{x} - I \mathbf{x})  \, , \, M_B =  d + e_4 ( \mathbf{r} + I \mathbf{r}) .
\ee
These last set of operators are special in that they lie in the even subalgebra and so describe pure rotations, that is, will rotate a unit four-vector to a unit four-vector.

\bibliography{quantum}

\begin{thebibliography}{9}
\expandafter\ifx\csname natexlab\endcsname\relax\def\natexlab#1{#1}\fi
\expandafter\ifx\csname bibnamefont\endcsname\relax
  \def\bibnamefont#1{#1}\fi
\expandafter\ifx\csname bibfnamefont\endcsname\relax
  \def\bibfnamefont#1{#1}\fi
\expandafter\ifx\csname citenamefont\endcsname\relax
  \def\citenamefont#1{#1}\fi
\expandafter\ifx\csname url\endcsname\relax
  \def\url#1{\texttt{#1}}\fi
\expandafter\ifx\csname urlprefix\endcsname\relax\def\urlprefix{URL }\fi
\providecommand{\bibinfo}[2]{#2}
\providecommand{\eprint}[2][]{\url{#2}}

\bibitem[{\citenamefont{Buhrman et~al.}(2012)\citenamefont{Buhrman, Christandl,
  and Schaffner}}]{Buhrman2012complete}
\bibinfo{author}{\bibfnamefont{H.}~\bibnamefont{Buhrman}},
  \bibinfo{author}{\bibfnamefont{M.}~\bibnamefont{Christandl}},
  \bibnamefont{and}
  \bibinfo{author}{\bibfnamefont{C.}~\bibnamefont{Schaffner}},
  \bibinfo{journal}{Phys. Rev. Lett.} \textbf{\bibinfo{volume}{109}},
  \bibinfo{pages}{160501} (\bibinfo{year}{2012}).

\bibitem[{\citenamefont{Lo et~al.}(2012)\citenamefont{Lo, Curty, and
  Qi}}]{lo2012measurement}
\bibinfo{author}{\bibfnamefont{H.}~\bibnamefont{Lo}},
  \bibinfo{author}{\bibfnamefont{M.}~\bibnamefont{Curty}}, \bibnamefont{and}
  \bibinfo{author}{\bibfnamefont{B.}~\bibnamefont{Qi}}, \bibinfo{journal}{Phys.
  Rev. Lett.} \textbf{\bibinfo{volume}{108}}, \bibinfo{pages}{130503}
  (\bibinfo{year}{2012}).

\bibitem[{\citenamefont{Nguimdo et~al.}(2011)\citenamefont{Nguimdo, Colet,
  Larger, and Pesquera}}]{nguimdo2011digital}
\bibinfo{author}{\bibfnamefont{R.}~\bibnamefont{Nguimdo}},
  \bibinfo{author}{\bibfnamefont{P.}~\bibnamefont{Colet}},
  \bibinfo{author}{\bibfnamefont{L.}~\bibnamefont{Larger}}, \bibnamefont{and}
  \bibinfo{author}{\bibfnamefont{L.}~\bibnamefont{Pesquera}},
  \bibinfo{journal}{Phys. Rev. Lett.} \textbf{\bibinfo{volume}{107}},
  \bibinfo{pages}{34103} (\bibinfo{year}{2011}).

\bibitem[{\citenamefont{Kanter et~al.}(2008)\citenamefont{Kanter, Kopelowitz,
  and Kinzel}}]{kanter2008public}
\bibinfo{author}{\bibfnamefont{I.}~\bibnamefont{Kanter}},
  \bibinfo{author}{\bibfnamefont{E.}~\bibnamefont{Kopelowitz}},
  \bibnamefont{and} \bibinfo{author}{\bibfnamefont{W.}~\bibnamefont{Kinzel}},
  \bibinfo{journal}{Phys. Rev. Lett.} \textbf{\bibinfo{volume}{101}},
  \bibinfo{pages}{84102} (\bibinfo{year}{2008}).

\bibitem[{\citenamefont{Kish and Sethuraman}(2004)}]{laszlo2004non}
\bibinfo{author}{\bibfnamefont{L.~B.} \bibnamefont{Kish}} \bibnamefont{and}
  \bibinfo{author}{\bibfnamefont{S.}~\bibnamefont{Sethuraman}},
  \bibinfo{journal}{Fluctuation and Noise Letters}
  \textbf{\bibinfo{volume}{4}}, \bibinfo{pages}{1} (\bibinfo{year}{2004}).

\bibitem[{\citenamefont{Kish et~al.}(2005)\citenamefont{Kish, Sethuraman, and
  Heszler}}]{kish2005}
\bibinfo{author}{\bibfnamefont{L.~B.} \bibnamefont{Kish}},
  \bibinfo{author}{\bibfnamefont{S.}~\bibnamefont{Sethuraman}},
  \bibnamefont{and} \bibinfo{author}{\bibfnamefont{P.}~\bibnamefont{Heszler}},
  \bibinfo{journal}{AIP Conference Proceedings} \textbf{\bibinfo{volume}{800}},
  \bibinfo{pages}{193} (\bibinfo{year}{2005}).

\bibitem[{\citenamefont{Klappenecker}(2004)}]{klappenecker2004remark}
\bibinfo{author}{\bibfnamefont{A.}~\bibnamefont{Klappenecker}},
  \bibinfo{journal}{Fluctuation and Noise Letters}
  \textbf{\bibinfo{volume}{4}}, \bibinfo{pages}{25} (\bibinfo{year}{2004}).

\bibitem[{\citenamefont{El-Hajjar et~al.}(2009)\citenamefont{El-Hajjar, Alamri,
  Wang, Zummo, and Hanzo}}]{Hajjar2009steered}
\bibinfo{author}{\bibfnamefont{M.}~\bibnamefont{El-Hajjar}},
  \bibinfo{author}{\bibfnamefont{O.}~\bibnamefont{Alamri}},
  \bibinfo{author}{\bibfnamefont{J.}~\bibnamefont{Wang}},
  \bibinfo{author}{\bibfnamefont{S.}~\bibnamefont{Zummo}}, \bibnamefont{and}
  \bibinfo{author}{\bibfnamefont{L.}~\bibnamefont{Hanzo}},
  \bibinfo{journal}{IEEE Trans. Wireless Comm.} \textbf{\bibinfo{volume}{8}},
  \bibinfo{pages}{3335} (\bibinfo{year}{2009}).

\bibitem[{\citenamefont{Doran and Lasenby}(2003)}]{Doran2003}
\bibinfo{author}{\bibfnamefont{C.~J.~L.} \bibnamefont{Doran}} \bibnamefont{and}
  \bibinfo{author}{\bibfnamefont{A.~N.} \bibnamefont{Lasenby}},
  \emph{\bibinfo{title}{Geometric Algebra for Physicists}}
  (\bibinfo{publisher}{Cambridge Univ Pr}, \bibinfo{address}{Cambridge},
  \bibinfo{year}{2003}).

\end{thebibliography}

\end{document}